\documentclass{tlp}
\submitted{4 February 2010}
\revised{30 April 2010}
\accepted{21 May 2010}

\usepackage{stmaryrd, amssymb, extarrows}
\usepackage{verbatim} 

\newtheorem{defn}{Definition}[section] 
\newtheorem{thm}{Theorem}[section] 
\newtheorem{prop}{Proposition}[section] 
\newtheorem{exmp}{Example}[section] 
\newtheorem{cor}{Corollary}[section] 

\newcommand{\nc}{\newcommand}

\nc{\Con}[1]{\mbox{Con}_{#1}}  

\nc{\pc}[3]{\mathsf{#1}_{#2}(#3)} 
\nc{\qval}[1]{\pc{qVal}{}{#1}} 
\nc{\qbound}[1]{\pc{qBound}{}{#1}} 

\newcommand{\schemenp}[1]{\mbox{#1}} 
\newcommand{\scheme}[2]{\schemenp{#1}(#2)} 

\newcommand{\clp}[1]{\scheme{CLP}{#1}} 

\newcommand{\sqclp}[3]{\scheme{SQCLP}{#1,#2,#3}} 

\newcommand{\cdom}{\mathcal{C}} 

\newcommand{\rdom}{\mathcal{R}} 
\newcommand{\hdom}{\mathcal{H}} 

\newcommand{\qdom}{\mathcal{D}} 
\newcommand{\aqdomd}[1]{D_{#1} \setminus \{\bt\}} 
\newcommand{\bqdomd}[1]{(D_{#1} \setminus \{\bt\}) \uplus \{?\}} 
\newcommand{\aqdom}{\aqdomd{}} 
\newcommand{\bqdom}{\bqdomd{}} 

\newcommand{\bt}{\mathrm{\mathbf{b}}} 
\newcommand{\tp}{\mathrm{\mathbf{t}}} 
\newcommand{\dleq}{\trianglelefteqslant} 
\newcommand{\dlt}{\vartriangleleft} 
\newcommand{\dgeq}{\trianglerighteqslant} 

\newcommand{\B}{\mathcal{B}} 
\newcommand{\U}{\mathcal{U}} 
\newcommand{\W}{\mathcal{W}} 

\newcommand{\simrel}{\mathcal{S}} 
\newcommand{\sid}{\simrel_{\mathrm{id}}} 

\newcommand{\toy}{\mathcal{TOY}}

\newcommand{\Prog}{\mathcal{P}} 

\newcommand{\Var}{\mathcal V\!ar} 
\newcommand{\War}{\mathcal W\!ar} 

\nc{\Exp}{\mbox{Exp}_{\bot}(\Sigma,B,\Var)} 
\nc{\TExp}{\mbox{Exp}(\Sigma,B,\Var)} 
\nc{\GExp}{\mbox{Exp}_{\bot}(\Sigma,B)} 
\nc{\TGExp}{\mbox{Exp}(\Sigma,B)} 
\nc{\Term}{\mbox{Term}_{\bot}(\Sigma,B,\Var)} 
\nc{\TTerm}{\mbox{Term}(\Sigma,B,\Var)} 
\nc{\GTerm}{\mbox{Term}_{\bot}(\Sigma,B)} 
\nc{\TGTerm}{\mbox{Term}(\Sigma,B)} 

\nc{\At}{\mbox{At}(\Sigma,B,\Var)} 
\nc{\GAt}{\mbox{GAt}(\Sigma,B)} 
\nc{\PAt}{\mbox{PAt}(\Sigma,B,\Var)} 
\nc{\GPAt}{\mbox{GPAt}(\Sigma,B)} 

\nc{\Atz}{\mbox{At}_{\Sigma}} 
\nc{\QAtz}{\mbox{At}_{\Sigma}(\qdom)} 

\nc{\sust}{\mbox{Subst}_\Sigma} 
\nc{\Sust}{\mbox{Subst}(\Sigma,B,\Var)} 
\nc{\GSust}{\mbox{GSubst}(\Sigma,B)} 

\nc{\Soln}[3]{\mbox{Sol}_{#1}^{#2}(#3)} 
\nc{\Sol}[2]{\Soln{#1}{}{#2}} 
\nc{\GSol}[2]{\mbox{GSol}_{#1}(#2)}
\nc{\Solc}[1]{\Sol{\cdom}{#1}}
\nc{\CAns}[2]{\mbox{C\!Ans}_{#1}(#2)} 

\newcommand{\qat}[2]{#1 \sharp #2} 
\newcommand{\cqat}[3]{\qat{#1}{#2} \Leftarrow #3} 

\newcommand{\qgets}[1]{\xleftarrow{#1}} 

\newcommand{\Int}[1]{\mbox{Int}_{#1}}
\newcommand{\Intdc}{\Int{\qdom,\cdom}} 

\newcommand{\ibot}{\bot\!\!\!\bot} 
\newcommand{\itop}{\top\!\!\!\top} 

\newcommand{\I}{\mathcal{I}} 

\nc{\closure}[1]{\mbox{cl}_{#1}} 

\newcommand{\Tp}{\mbox{T}_{\!\Prog}} 

\newcommand{\model}[1]{~{\models_{#1}}~} 

\newcommand{\M}[1]{\mathcal{M}_{#1}} 
\newcommand{\Mp}{\M{\Prog}} 

\newcommand{\diff}{~{\Longleftrightarrow_{\mathrm{def}}}~} 
\newcommand{\eqdef}{~{=_{\mathrm{def}}}~} 
\newcommand{\union}{\bigcup} 
\newcommand{\inter}{\bigcap} 
\newcommand{\supr}{\bigsqcup} 
\newcommand{\infi}{\bigsqcap} 
\newcommand{\entail}[1]{~{\succcurlyeq_{#1}}~} 

\newcommand{\sep}{\talloblong} 

\newcommand{\NAT}{\mathbb{N}}
\newcommand{\REAL}{\mathbb{R}}

\newcommand{\tup}[1]{\overline{#1}}   
\newcommand{\ntup}[2]{\tup{#1}_{#2}}  

\newcommand{\infx}[2]{\ {\vdash}_{\!#1}^{\!#2}\ } 
\newcommand{\infxx}[2]{\ {\vdash\!\!\vdash}_{\!#1}^{\!#2}\ } 

\newcommand{\SQCHL}{\mbox{SQCHL}} 
\newcommand{\sqchln}[4]{\infx{#1,#2,#3}{#4}} 
\newcommand{\sqchlrdc}{\sqchln{\simrel}{\qdom}{\cdom}{}} 
\newcommand{\sqchlrdcn}[1]{\sqchln{\simrel}{\qdom}{\cdom}{#1}} 

\newcommand{\isqchln}[4]{\infxx{#1,#2,#3}{#4}} 
\newcommand{\isqchlrdc}{\isqchln{\simrel}{\qdom}{\cdom}{}} 

\begin{document}

\long\def\comment#1{}

\title[A Declarative Semantics for CLP with Qualification and Proximity]
{A Declarative Semantics for CLP\\ with Qualification and Proximity \thanks{This work has been partially supported by the Spanish projects STAMP (TIN2008-06622-C03-01), PROMETIDOS--CM (S2009TIC-1465) and GPD--UCM (UCM--BSCH--GR58/08-910502).}}

\author[M. Rodr\'iguez-Artalejo and C. A. Romero-D\'iaz]
{MARIO RODR\'IGUEZ-ARTALEJO and CARLOS A. ROMERO-D\'IAZ \\
Departamento de Sistemas Inform\'aticos y Computaci\'on, Universidad Complutense\\
Facultad de Inform\'atica, 28040 Madrid, Spain\\
\email{mario@sip.ucm.es, cromdia@fdi.ucm.es}
}

\pagerange{\pageref{firstpage}--\pageref{lastpage}}
\volume{\textbf{10} (3):}
\jdate{March 2002}
\setcounter{page}{1}
\pubyear{2002}

\maketitle


\begin{abstract}
Uncertainty in Logic Programming has been investigated during the last decades,
dealing with various extensions of the classical LP paradigm and different applications.
Existing proposals rely on different approaches, such as 
clause annotations based on uncertain truth values,
qualification values as a generalization of uncertain truth values,
and unification based on proximity  relations.
On the other hand,  the CLP scheme 
has established itself as a powerful extension of LP
that supports efficient computation over specialized domains 
while keeping a  clean declarative semantics.
In this paper we propose a new scheme SQCLP
designed as an extension of CLP that supports qualification values and proximity relations.
We show that several  previous proposals 
can be viewed as particular cases of the new scheme, obtained by partial instantiation.
We present a declarative semantics for SQCLP that is based on observables, 
providing fixpoint and proof-theoretical characterizations of least program models
as well as an implementation-independent  notion of goal solutions.
\end{abstract}


\section{Introduction}
\label{sec:introduction}


Many extensions of logic programming (shortly LP) to deal with uncertainty have been proposed in the last decades.
A line of research not related to this paper is based on probabilistic extensions of LP such as \cite{NS92}.
Other proposals in the field replace classical two-valued logic 
by some kind of many-valued logic whose truth values 
can be attached to computed answers and are usually  interpreted as certainty degrees.
The next paragraphs summarize some relevant approaches of this kind. 

There are extensions of  LP using annotations in program clauses
to compute a certainty degree for the head atom from the certainty degrees previously computed for the body atoms.
This line of research includes the seminal proposal of Quantitative Logic Programming  by \cite{VE86} 
and inspired later works such as the Generalized Annotated logic Programs (shortly GAP) by  \cite{KS92}
and  the QLP scheme for Qualified LP \cite{RR08}.
While \cite{VE86} and other early approaches used real numbers of the interval $[0,1]$ as certainty degrees,
QLP and GAP take elements from a parametrically given lattice to be used in annotations and attached to computed answers. 
In the case of QLP, the lattice is called a {\em qualification domain} 
and its elements (called {\em qualification values})  are not always understood as certainty degrees.
As argued in  \cite{RR08}, GAP is a more general framework, 
but QLP's semantics  have some advantages for its intended scope.

There are also extended LP languages based on fuzzy logic 
which can be classified into two major lines.
The first line includes Fuzzy LP languages such as \cite{Voj01,GMV04} 
and the Multi-Adjoint LP (shortly MALP) framework by \cite{MOV01a}.
All these approaches extend classical LP by using clause annotations and a fuzzy interpretation of the connectives
and aggregation operators occurring in program clauses and goals. 
There is a relationship between Fuzzy LP and GAP that has been investigated in \cite{KLV04}.
Intended applications of Fuzzy LP languages include expert knowledge representation. 

The second line includes Similarity-based LP (shortly SLP)
 in the sense of \cite{Ses02} and related proposals,
which keep the classical syntax of LP clauses but use a {\em similarity relation} over a set of symbols $S$ 
to allow ``flexible'' unification of syntactically different symbols with a certain approximation degree.
Similarity relations over a given set $S$ have been defined in \cite{Zad71,Ses02} and related literature as 
fuzzy relations represented by mappings $\simrel : S \times S \to [0,1]$ which satisfy reflexivity, symmetry and transitivity  axioms analogous to those required for classical equivalence relations. A more general notion called {\em proximity relation} was 
introduced in \cite{DP80} by omitting the transitivity axiom. 
As  noted by  \cite{SM99} and other authors, the transitivity property required for similarity relations
may conflict with user's intentions in some cases. 
The \textsf{Bousi}$\sim$\textsf{Prolog} language \cite{JR09} 
has been designed with the aim of generalizing SLP to work with proximity relations.
A different generalization of SLP is the SQLP scheme \cite{CRR08},
designed as an extension of the QLP scheme.
In addition to clause annotations in QLP style, SQLP uses a given similarity 
relation $\simrel : S \times S \to D$ (where $D$ is the carrier set of a parametrically given qualification domain)
in order to support flexible unification.
In the sequel we use the acronym SLP as including proximity-based LP languages also.
Intended applications of SLP include flexible query answering.  
An analogy of proximity relations in a different context (namely partial constraint satisfaction) can be found in \cite{FW92}, where several metrics are proposed to measure the proximity between the solution sets of two different constraint satisfaction problems.

Several of the above mentioned LP extensions (including GAP, QLP, the Fuzzy LP language in  \cite{GMV04} and SQLP)
 have used constraint solving as an implementation technique.
However, we only know two approaches which have been conceived as extensions of the classical CLP scheme \cite{JL87}.
Firstly, \cite{Rie98phd} extended the formulation of CLP by \cite{HS88} with quantitative LP in the sense of \cite{VE86};
this work was  motivated by problems from the field of natural language processing.
Secondly, \cite{BMR01} proposed  a semiring-based approach to CLP, 
where constraints are solved in a soft way with levels of consistency represented by values of a semiring.
This approach was motivated by constraint satisfaction problems and implemented with {\tt clp(FD,S)} in \cite{GC98} for a particular class of semirings which enable to use local consistency algorithms.
The relationships between \cite{Rie98phd,BMR01} and the results of this paper will be further discussed in Section \ref{sec:conclusions}.

Finally, there are a few preliminary attempts to combine some of the above mentioned approaches 
with the Functional Logic Programming (shortly FLP) paradigm 
found in languages such as {\sf Curry} \cite{curry} and $\toy$ \cite{toy}. 
Similarity-based unification for FLP languages has been investigated by \cite{MP07}, 
while \cite{CRR09} have proposed a generic scheme QCFLP designed as a common extension of
the two schemes CLP and QLP with first-order FLP features.


In this paper we propose a new extension of CLP that supports qualification values and proximity relations.
More precisely, we define a generic scheme SQCLP whose instances $\sqclp{\simrel}{\qdom}{\cdom}$ are
parameterized by a proximity relation $\simrel$, a qualification domain $\qdom$ and a constraint domain $\cdom$. 
We will show that several previous proposals 
can be viewed as particular cases of SQCLP, obtained by partial instantiation.
Moreover, we will present a declarative semantics for SQCLP that is inspired in the observable CLP semantics by \cite{GDL95}
and provides fixpoint and proof-theoretical characterizations of least program models
as well as an implementation-independent notion of goal solution that can be used to specify the expected behavior of goal solving systems.

The reader is assumed to be familiar with the semantic foundations of LP and CLP.
The rest of the paper is structured as follows:
Section \ref{sec:cbasis} introduces constraint domains, qualification domains and proximity relations.
Section \ref{sec:sqclp} presents the \mbox{SQCLP} scheme and the main results on its declarative semantics.
Finally, Section \ref{sec:conclusions} concludes by  giving a discussion of related approaches (many of which can be viewed as particular cases of \mbox{SQCLP}) and pointing to some lines open for future work. 
Due to space limits, we have preferred to include examples rather than proofs.
A widely extended version including detailed proofs is available as Technical Report \cite{RR10TR}.

\section{Computational Basis}
\label{sec:cbasis}

\subsection{Constraint Domains}
\label{sec:cbasis:cdoms}


As in the CLP Scheme, we will work with constraint domains related to signatures.
We assume a {\em universal programming signature} $\Gamma = \langle DC, DP
\rangle$ where $DC = \bigcup_{n \in \NAT}  DC^n$ and $DP = \bigcup_{n \in \NAT}  DP^n$ are
countably infinite and mutually disjoint sets of free function symbols (called {\em data constructors} in
the sequel) and {\em defined predicate} symbols, respectively, ranked by arities. 
We will use {\em domain specific signatures}
$\Sigma = \langle DC, DP, PP \rangle$ extending $\Gamma$ with a disjoint set $PP = \bigcup_{n \in
\NAT}  PP^n$ of {\em primitive predicate} symbols, also ranked by arities. The idea is that
primitive predicates come along with constraint domains, while defined predicates are specified in
user programs. Each $PP^n$ may be any countable set of $n$-ary predicate symbols.


{\em Terms} have the syntax $t ::= X|u|c(\ntup{t}{n})$, where $X \in \Var$, $u \in B$ and $c \in DC^n$,
assuming a countably infinite set of variables $\Var$ and a set of basic values $B$ and using $\ntup{t}{n}$ as
a shorthand for $t_1, \ldots, t_n$.
The set of ground terms is noted $\TGTerm$.
As usual, {\em substitutions}  are defined as mappings $\sigma$ assigning terms
to variables and extended to act over other syntactic objects $o$ in the natural way.
The result of applying $\sigma$ to $o$ is noted as $o\sigma$.


Several formal notions of {\em constraint domain} are known in the literature. In this paper, 
constraint domains of signature $\Sigma$ are relational structures of the
form $\cdom = \langle C, \{p^\cdom \mid p \in PP\}\rangle$ consisting of a carrier set $C = \TGTerm$ and
an interpretation $p^{\cdom} : C^n \to \{0,1\}$ for each $p \in PP^n$.
For the examples in this paper we will use the {\em real constraint domain} $\rdom$ well known as the basis of
the $\clp{\rdom}$ language and system \cite{JMSY92}. In our setting we represent $\rdom$ with set of basic
values $B = \REAL$ and primitive predicates $op_{+}, op_{\times}, \ldots \in PP^3$ and $cp_{>}, cp_{\geq}, \ldots \in PP^2$
defined to represent the usual arithmetic and comparison operations over $\REAL$. Other useful constraint domains are: the {\em Herbrand} domain $\mathcal{H}$, intended to work just with equality constraints; and $\mathcal{FD}$, intended to work with constraints involving integer values and finite domain variables.


Given a constraint domain $\cdom$, we will work with {\em atoms} of three kinds: defined atoms $A :
r(\ntup{t}{n})$, where $r \in DP^n$ and $t_i$ are terms; primitive atoms $\kappa : p(\ntup{t}{n})$, 
where $p \in PP^n$ and $t_i$ are terms; and equations $t$ {\sf ==} $s$, where $t, s$ are terms 
and {\sf ==} is the {\em equality symbol}. 
Primitive atoms and equations are called {\em atomic $\cdom$-constraints}. More
generally, $\cdom$-constraints $\pi$ are built from atomic $\cdom$-constraints using logical conjunction $\land$,
existential quantification $\exists$, and sometimes other logical operations.
Constraints of the form $\exists X_1 \ldots \exists X_n(B_1 \land \ldots \land B_m)$ --where $B_j ~ (1 \leq j \leq m)$ are atomic-- are called {\em existential}.


The set of all $\cdom$-constraints is noted $\Con{\cdom}$.
Constraints are interpreted by means of {\em $\cdom$-valuations} $\eta \in \mbox{Val}_{\cdom}$,
which are ground substitutions.
The set $\Solc{\Pi}$ of solutions of $\Pi \subseteq \Con{\cdom}$ includes all the valuations $\eta$ such that $\Pi\eta$
is true when interpreted in $\cdom$.
$\Pi \subseteq \Con{\cdom}$ is called {\em satisfiable} if $\Solc{\Pi} \neq \emptyset$ and {\em unsatisfiable} otherwise.
$\pi \in \Con{\cdom}$ {\em is entailed} by $\Pi \subseteq \Con{\cdom}$ 
(noted  $\Pi \model{\cdom} \pi$) iff $\Solc{\Pi} \subseteq \Solc{\pi}$.

As a simple illustration consider $\Pi = \{cp_{\geq}(A,3.0),\, op_{+}(A,A,X), op_{\times}(2.0,A,Y)\}$
$\subseteq \Con{\rdom}$. Clearly, $\eta \in \Sol{\rdom}{\Pi}$ holds iff $\eta(A)$,
$\eta(X)$ and $\eta(Y)$ are real numbers $a, x, y \in \REAL$ such that $a \geq 3.0$, $a+a = x$ and
$2.0\times a = y$. Then $\Sol{\rdom}{\Pi} \subseteq \Sol{\rdom}{X == Y}$.
Therefore, assuming $c \in DC^1$ one has $\Pi \model{\rdom} c(X)  ==  c(Y)$.

\subsection{Qualification Domains}
\label{sec:cbasis:qdoms}

As mentioned in the Introduction, qualification values were introduced as a generalization
of certainty values in \cite{RR08}. They are elements of special lattices called {\em qualification
domains} and defined as structures $\qdom = \langle D, \dleq, \bt, \tp, \circ \rangle$ verifying the following requirements:
\begin{enumerate}
\item
$\langle D, \dleq, \bt, \tp \rangle$ is a lattice with extreme points $\bt$ (called {\em infimum} or {\em bottom} element) and $\tp$ (called {\em maximum} or {\em top} element) w.r.t. the partial ordering $\dleq$ (called {\em qualification ordering}). For given elements  $d, e \in D$, we  write $d \sqcap e$ for the {\em greatest lower bound} ($glb$) of $d$ and $e$, and $d \sqcup e$ for the {\em least upper bound} ($lub$) of $d$ and $e$. We also write $d \dlt e$ as abbreviation for $d \dleq e \land d \neq e$.
    \item $\circ : D \times D \rightarrow D$, called {\em attenuation operation}, verifies the following axioms:
        \begin{enumerate}
            \item $\circ$ is associative, commutative and monotonic w.r.t. $\dleq$.
            \item $\forall d \in D : d \circ \tp = d$ and $d \circ \bt = \bt$.
            \item $\forall d, e \in D  : d \circ e \dleq e$ and even $\bt \neq d \circ e \dleq e$ if $d,e \in \aqdom$.
            \item $\forall d, e_1, e_2 \in D : d \circ (e_1 \sqcap e_2) = (d \circ e_1) \sqcap (d \circ e_2)$.
        \end{enumerate}
\end{enumerate}

Actually, axioms (2)(b.2) and (2)(c.1) are redundant because they can be derived  from the
other axioms.\footnote{The authors are thankful to G. Gerla for pointing out this fact.}
For any $S = \{e_1, e_2, \ldots, e_n\} \subseteq D$, the $glb$ (also called {\em infimum} of $S$)
exists and can be computed as $\infi S =e_1 \sqcap e_2 \sqcap \cdots \sqcap e_n$ 
(which reduces to $\tp$ in the case $n = 0$).
The dual claim concerning $lub$s is also true. 
As an easy consequence of the axioms, one gets the identity $d \circ \infi S =  \infi \{d \circ e \mid e \in S\}$.

The following basic qualification domains were also introduced in \cite{RR08}.

\smallskip
\noindent
\textbf{The Domain of Classical Boolean Values} is $\B \eqdef \langle \{0,1\}, \leq, 0, 1, \land \rangle$,
where $0$ and $1$ stand for the two classical truth values  \emph{false} and \emph{true}, $\leq$ is
the usual numerical ordering over $\{0,1\}$, and $\land$ stands for the classical conjunction operation
over $\{0,1\}$.

\smallskip
\noindent
\textbf{The Domain of Uncertainty Values}
is $\U \eqdef \langle \mbox{U}, \leq, 0, 1,\times \rangle$, where $\mbox{U} = [0,1] =
\{d \in \REAL \mid 0 \le d \le 1\}$, $\le$ is the usual numerical ordering, and $\times$ is the
multiplication operation. The top element $\tp$ is $1$ and for any finite $S \subseteq \mbox{U}$
one has $\infi S = \mbox{min}(S)$, which is $1$ if $S = \emptyset$.
Elements of $\U$ are intended to represent certainty degrees.

\smallskip
\noindent
\textbf{The Domain of Weight Values}
is $\W \eqdef \langle \mbox{P}, \ge, \infty, 0, + \rangle $, where $\mbox{P} =
[0,\infty] = \{d \in \REAL \cup \{\infty\} \mid d \ge 0\}$, $\geq$ is the reverse of the usual
numerical ordering (with $\infty \ge d$ for any $d \in \mbox{P}$), and $+$ is the addition
operation (with $\infty + d = d + \infty = \infty$ for any $d \in \mbox{P}$). The top element $\tp$
is $0$ and for any finite $S \subseteq \mbox{P}$ one has $\infi S = \mbox{max}(S)$, which is $0$ if $S = \emptyset$.
Elements of $\W$ are intended to represent proof costs, measured as the weighted depth of proof trees.

\smallskip
Given qualification domains $\qdom_1$ and $\qdom_2$, their
{\em strict cartesian product} $\qdom_1 \otimes \qdom_2$
is $\qdom \eqdef \langle D, \dleq, \bt, \tp, \circ \rangle$ where
$D  = D_1 \otimes D_2 \eqdef ((D_1 \!\setminus\! \{\bt_1\}) \times (D_2 \!\setminus\! \{\bt_2\})) \cup \{(\bt_1, \bt_2)\}$,
the partial ordering $\dleq$ is defined as $(d_1,d_2) \dleq (e_1,e_2) \diff d_1 \dleq_1 e_1$ and $d_2 \dleq_2 e_2$,
and the attenuation operator $\circ$ is defined as $(d_1,d_2) \circ (e_1,e_2) \eqdef (d_1 \circ_1 e_1, d_2 \circ_2 e_2)$.
It can be proved that $\qdom_1 \otimes \qdom_2$ is again a qualification domain.\footnote{This result
refines a similar one for ordinary cartesian products presented in \cite{RR08}.}

In Section \ref{sec:sqclp} we will need the following definition, that refines a similar one given in \cite{CRR09}.

\begin{defn} [Expressing $\qdom$ in $\cdom$]
\label{dfn:expressible} A qualification domain $\qdom$ is expressible in a constraint domain $\cdom$
if there is an injective embedding mapping $\imath : \aqdom \to C$ and moreover:
  \begin{enumerate}
     \item
     There is a $\cdom$-constraint $\qval{X}$ such that      
     $\Solc{\qval{X}}$ is the set of all $\eta \in \mbox{Val}_\cdom$  verifying  $\eta(X) \in ran(\imath)$. 
     \item
    There is a $\cdom$-constraint $\qbound{X,Y,Z}$ encoding ``$x \dleq y \circ z$'' in the following sense: 
    any $\eta \in \mbox{Val}_\cdom$ such that $\eta(X) = \iota(x)$, $\eta(Y) = \iota(y)$ and $\eta(Z) = \iota(z)$
    verifies $\eta \in \Solc{\qbound{X,Y,Z}}$ iff $x \dleq y \circ z$. 
  \end{enumerate}
In addition, if $\qval{X}$ and $\qbound{X,Y,Z}$ can be chosen as existential constraints, we say that $\qdom$ is 
{\em existentially expressible} in $\cdom$. \mathproofbox
\end{defn}

We can prove that  $\B$, $\U$, $\W$ and any qualification domain built from these with the help of $\otimes$ is existentially 
expressible in any constraint domain $\cdom$ that includes the basic values and computational features of $\rdom$.
For instance,  $\U \otimes \W$ can be expressed in $\rdom$ using a binary data constructor {\sf pair} 
$\in DC^2$ and taking: $\iota(x,y) \eqdef {\sf pair}(x,y)$; 
$\qval{X} : \exists X_1 \exists X_2 (X == {\sf pair}(X_1,X_2) \land cp_{<}(0,X_1) \land cp_{\leq}(X_1,1)   \land cp_{\leq}(0,X_2))$; 
and $\qbound{X,Y,Z}$ built in a suitable way. The interested reader is referred to \cite{RR10TR} for other examples of qualification domains which can be existentially expressed in $\mathcal{FD}$.

\subsection{Proximity Relations}
\label{sec:domains:simrels}

Similarity and proximity relations have been introduced in Section \ref{sec:introduction}. In the rest of this
paper we will focus on triples $\langle \simrel, \qdom, \cdom \rangle$ fulfilling the following requirements:

\begin{defn}[Admissible triples]
\label{defn:simrel:admissible}
An {\em admissible triple} $\langle \simrel, \qdom, \cdom \rangle$ consist of a constraint domain $\cdom$ with
signature $\Sigma = \langle DC, DP, PP \rangle$ and set of basic values $B$, a qualification domain $\qdom$
expressible in $\cdom$ and a mapping $\simrel : S \times S \to D$ satisfying the following properties:
\begin{enumerate}
\item
$S = \Var \uplus B \uplus DC \uplus DP \uplus PP$.
\item
$\simrel$ is a {\em $\qdom$-valued proximity relation} such that $\simrel(x,x) = \tp$ (reflexivity)
and $\simrel(x,y) = \simrel(y,x)$ (symmetry) hold
for all $x,y \in S$. In the case that $\simrel$ satifies also $\simrel(x,z) \dgeq \simrel(x,y) \sqcap \simrel(y,z)$ (transitivity) for all
$x,y,z \in S$, it is called {\em $\qdom$-valued similarity relation}.
\item
$\simrel$ restricted to $\Var$ behaves as the identity,
i.e. $\simrel(X,X) = \tp$ for all $X \in \Var$ and $\simrel(X,Y) = \bt$ for all $X, Y \in \Var$ such that  $X \neq Y$.
\item
For any  $x, y \in S$, $\simrel(x,y) \neq \bt$ can happen only if:
(a) $x = y$ are identical; or else
(b) $x,y \in B$ are basic values; or else
(c) $x,y \in DC$ are data constructor symbols with the same arity; or else
(d) $x,y \in DP$ are defined predicate symbols with the same arity. \mathproofbox
\end{enumerate}
\end{defn}


$\qdom$-valued proximity relations generalize the $\qdom$-valued similarity relations first introduced  in \cite{CRR08}.
When $\qdom$ is chosen as  the qualification domain $\U$,  the previous
definition provides proximity and similarity relations in the sense of \cite{Zad71,DP80}. 
In this case, a proximity degree $\simrel(x,y) = d \in [0,1]$ can be naturally interpreted 
as a {\em certainty degree} for the assertion that $x$ and $y$ are interchangeable. 
On the other hand,  if $\simrel$ is $\W$-valued, then $\simrel(x,y) = d \in [0,\infty]$ can be interpreted 
as a {\em cost} to be paid for $y$ to play the role of $x$.


As mentioned in the Introduction, the transitivity property postulated for similarity relations 
may be counterintuitive in some cases. For instance, assume nullary constructors 
\texttt{colt}, \texttt{cold} and \texttt{gold} intended to represent
words composed of four letters. Then, measuring the proximity between such words might reasonably
lead to a $\U$-valued proximity relation $\simrel$ such that  $\simrel(\texttt{colt},\texttt{cold})
= 0.9$, $\simrel(\texttt{cold},\texttt{gold}) = 0.9$ and $\simrel(\texttt{colt},\texttt{gold}) =
0.4$. On the other hand, insisting on  $\simrel$ to  be transitive would enforce the unreasonable
condition $\simrel(\texttt{colt},\texttt{gold}) \geq 0.9$. Therefore, a similarity relation would not be appropriate in
this case.  


The special mapping $\sid : S \times S \to D$ defined as $\sid(x,x) = \tp$ for all $x \in S$ and
$\sid(x,y) = \bt$ for all $x,y \in S$, $x \neq y$ is trivially a $\qdom$-valued similarity (and
therefore, also a proximity) relation called the \emph{identity}.


In the rest of this paper, the notations $\simrel$, $\qdom$ and $\cdom$ are always understood as
the components of some given admissible triple
and the proximity relation $\simrel$ is not required to be transitive. 
As noted in \cite{Ses02} and related works, $\simrel$ can be naturally extended to act over terms. 
The extension, also noted $\simrel$, works as specified by the recursive equations displayed below:

\begin{itemize}
\item
$\simrel(t,t) = \tp$
for every term $t$. 
\item
$\simrel(X,t) = \simrel(t,X) = \bt$
for $X \in \Var$ and for any term $t \neq X$.
\item
$\simrel(c(\ntup{t}{n}), c'(\ntup{t'}{n}) = \bt$
for $c \in DC^n, c' \in DC^m$ with $n \neq m$.
\item
$\simrel(c(\ntup{t}{n}), c'(\ntup{t'}{n})) = \simrel(c,c') \sqcap \simrel(t_1,t_1') \sqcap \ldots \sqcap \simrel(t_n,t_n')$
for $c, c' \in DC^n$.
\end{itemize}

Analogously, $\simrel$ can be extended to work over atoms and other syntactic objects.
The following definition combines $\simrel$ with constraint entailment, 
leading to a kind of relations over terms which will play a crucial role for the semantics of equations in SQCLP.


\begin{defn}[Constraint-based term proximity at level $\lambda$]
\label{defn:Pi-prox} 
Assume  $\lambda \in
\aqdom$ and $\Pi \subseteq \Con{\cdom}$. We will say that two terms $t$ and $s$ are {\em $\simrel$-close at
level $\lambda$ w.r.t. $\Pi$} (in symbols, $t  \approx_{\lambda, \Pi} \!s$) iff  there are two terms
$\hat{t}$, $\hat{s}$ such that $\Pi \model{\cdom} t == \hat{t}$, $\Pi \model{\cdom} s == \hat{s}$ and $\simrel(\hat{t}
,\hat{s}) \dgeq \lambda$. \mathproofbox
\end{defn}

It can be proved that $\approx_{\lambda,\Pi}$ is a reflexive and symmetric relation over the set of terms,
that is even transitive in case that $\simrel$ is a similarity relation. As a simple example, assume 
$\qdom = \U$, $\cdom = \rdom$ and  $\simrel$ such that $\simrel(c',c) = \simrel (c,c'') = 0.8$ and $\simrel(c',c'') =  0.6$
for some $c, c' , c'' \in DC^2$. Let $\Pi = \{op_{+}(A,A,X), op_{\times}(2.0,A,Y),$ $Z == c(X,Y)\} \subseteq \Con{\rdom}$.
Note that this choice of $\Pi$ ensures $\Pi \model{\rdom} X == Y$. 
Then $c'(Y,X) \approx_{0.7, \Pi} Z$ holds, because $\Pi \model{\rdom} c'(Y,X) == c'(X,X)$, 
$\Pi \model{\rdom} Z == c(X,X)$ and $\simrel(c'(X,X),c(X,X)) = 0.8 \geq 0.7$.

\section{The SQCLP Scheme}
\label{sec:sqclp}

\subsection{Programs, Interpretations and Models}
\label{sec:sqclp:programs}


The scheme SQCLP has instances $\sqclp{\simrel}{\qdom}{\cdom}$ where $\langle \simrel,\qdom,\cdom \rangle$
is an admissible triple. 
A $\sqclp{\simrel}{\qdom}{\cdom}$-program is a set $\Prog$ of  \emph{qualified 
program rules} (also called \emph{qualified clauses}) 
$C : A \qgets{\alpha} \qat{B_1}{w_1}, \ldots, \qat{B_m}{w_m}$, where $A$ is a defined atom, 
$\alpha \in \aqdom$ is  called the   {\em attenuation factor} of the clause and 
each $\qat{B_j}{w_j} ~ (1 \le j \le m)$ is an atom $B_j$
annotated with a so-called {\em threshold value} $w_j \in \bqdom$.
The intended meaning of $C$ is as follows:  
if for all $1 \leq j \leq m$ one has $\qat{B_j}{e_j}$ (meaning that $B_j$ holds with qualification value $e_j$)
for some $e_j \dgeq^? w_j$,
then $\qat{A}{d}$ (meaning that $A$ holds with qualification value $d$)
can be inferred for any $d \in \aqdom$ such that $d \dleq \alpha \circ \infi_{j = 1}^m e_j$. 
By convention, $e_j \dgeq^? w_j$ means $e_j \dgeq w_j$ if $w_j ~{\neq}~?$ and is identically true otherwise.
In practice threshold values equal to `?' and attenuation values equal to $\tp$ can be omitted. 


As motivating example, consider a $\sqclp{\simrel}{\U\!\!\otimes\!\!\W}{\rdom}$-program $\Prog$ 
including the clauses and equations for $\simrel$ displayed in Figure \ref{fig:sample}. 
From Subsection \ref{sec:cbasis:qdoms} recall that qualification values in $\U\!\otimes\!\W$ are pairs
$(d,e)$ (where $d$ represents a certainty degree and $e$ represents a proof cost),
as well as the behavior of $\dleq$ and $\circ$ in $\U\!\otimes\!\W$. 
Consider the problem of proving $\qat{\texttt{goodWork(king\_liar)}}{(d,e)}$ from $\Prog$.
This can be achieved for $d = 0.75 \times \mbox{min}\{d_1,d_2\}$, $e = 3 + \mbox{max}\{e_1,e_2\}$
by using $R_1$ instantiated by $\{{\tt X} \mapsto {\tt king\_liar}, {\tt Y} \mapsto {\tt shakespeare}\}$,
and going on to prove  $\qat{\texttt{famousAuthor(shakespeare)}}{(d_1,e_1)}$
for some  $d_1 \geq 0.5$,  $e_1 \leq 100$ 
and $\qat{\texttt{wrote(shakespeare,king\_liar)}}{(d_2,e_2)}$ for some $d_2$, $e_2$.
Thanks to $R_2$, $R_3$ and $\simrel$, these proofs succeed with $(d_1,e_1) = (0.9,1)$ and $(d_2,e_2) = (0.8,2)$.
Therefore, the  desired proof succeeds with certainty degree  $d = 0.75 \times \mbox{min}\{0.9,0.8\} = 0.6$, 
and proof cost $e = 3 + \mbox{max}\{1,2\} = 5$.

\begin{figure}[h]
\figrule
\small
\flushleft
\hspace{3mm} $R_1$ : \verb+goodWork(X) <-(0.75,3)- famousAuthor(Y)#(0.5,100), wrote(Y,X)#?+\\
\hspace{3mm} $R_2$ : \verb+famousAuthor(shakespeare) <-(0.9,1)-+\\
\hspace{3mm} $R_3$ : \verb+wrote(shakespeare,king_lear) <-(1,1)-+\\[2mm]
\hspace{1cm} $\simrel$\verb+(king_lear,king_liar) = (0.8,2)+
\caption{$\sqclp{\simrel}{\,\U\!\otimes\!\W}{\rdom}$ Program Fragment}
\label{fig:sample}
\figrule
\vspace{-3mm}
\end{figure}


The more technical $\sqclp{\simrel}{\U}{\rdom}$-program $\Prog$ presented below will serve as a
\emph{running example} in the rest of the paper.

\begin{exmp}[Running example]
\label{exmp:running} 
Assume $c, c' \in DC^1$, $p, p'\!, q \in DP^2$, $r \in DP^3$ and
$\simrel$ such that $\simrel(c,c') = 0.9$ and $\simrel(p,p') = 0.8$. Let
$\Prog$ consist of the following program rules:\\[1mm]
\begin{tabular}{ll}
\small $R_1 : q(X,c(X)) \qgets{1.0}$ & ~\, \small $R_3 : r(c(X),Y,Z) \qgets{0.9} \qat{q(X,Y)}{0.8}, \qat{cp_{\geq}(X,0.0)}{?}$ \\
\small $R_2 : p(c(X),Y) \qgets{0.9} \qat{q(X,Y)}{0.8} $ & \mathproofbox \\
\end{tabular}
\end{exmp}


The declarative semantics for SQCLP presented in the rest of this section is inspired 
by the $\mathcal{S}_{2}$ semantics for CLP given in \cite{GDL95}. 
We use {\em qualified constrained atoms} (or simply {\em qc-atoms}) of the form $\cqat{A}{d}{\Pi}$, 
intended to assert that the validity of atom  $A$ with qualification degree $d \in D$
is entailed by the constraint set $\Pi \subseteq \Con{\cdom}$. 
A qc-atom is called {\em defined}, {\em primitive} or {\em equational} according to the syntactic form of $A$; 
and it is called {\em observable} iff $d \in \aqdom$ and $\Pi$ is satisfiable. 
In the sequel we restrict our attention to observable qc-atoms, 
viewing them as observations of computed answers for atomic goals.
We use an {\em entailment relation} $\entail{\qdom,\cdom}$ to capture some implications between
qc-atoms whose validity depends neither on the proximity relation $\simrel$ nor on program clauses.\footnote{In 
\cite{CRR08} we used a different entailment relation that depends on $\simrel$
and does not work properly if $\simrel$ is not transitive.}
Formally, given $\varphi : \cqat{A}{d}{\Pi}$ and $\varphi' : \cqat{A'}{d'}{\Pi'}$,
we say that $\varphi$ $(\qdom, \cdom)$-\emph{entails} $\varphi'$ (in symbols, $\varphi \entail{\qdom,\cdom} \varphi'$)
iff there is some substitution $\theta$ such that $A' = A\theta$, $d' \dleq d$ and $\Pi' \model{\cdom} \Pi\theta$.
The example below illustrates these notions:

\begin{exmp}[Observable qc-atoms and $(\qdom, \cdom)$-entailment]
\label{exmp:qc-atoms} Building upon Example \ref{exmp:running}, let
$\Pi = \{cp_{>}(X,1.0), ~ op_{+}(A,A,X), ~ op_{\times}(2.0,A,$ $Y)\}$ and
$\Pi' = \{cp_{\geq}(A,3.0), ~ op_{\times}(2.0,$ $A,X), ~ op_{+}(A,A,Y)\}$.
Then, the following are observable qc-atoms: \\[2mm]
\begin{tabular}{l@{\hspace{1cm}}l}
\small $\varphi_1 : \cqat{q(X,c'(Y))}{0.9}{\Pi}$ 
& \small $\varphi_3 : \cqat{r(c'(Y),c(X),Z)}{0.8}{\Pi}$ \\
\small $\varphi_2 : \cqat{p'(c'(Y),c(X))}{0.8}{\Pi}$ 
& \small $\varphi'_3 : \cqat{r(c'(Y),c(X),c(Z'))}{0.7}{\Pi'}$ \\
\end{tabular}\\[2mm]
and $\varphi_3 \entail{\U, \rdom} \varphi'_3$ holds,
since $\theta = \{Z \mapsto c(Z')\}$ verifies $r(c'(Y),c(X),$ $c(Z')) = r(c'(Y),c(X),Z)\theta$, $0.7 \leq 0.8$
and $\Pi' \model{\rdom} \Pi\theta$. \mathproofbox
\end{exmp}


The intended meaning of $\!\!\entail{\qdom,\cdom}\!\!$ motivates the first sentence in the next definition.

\begin{defn}[Interpretations]
\label{defn:interpretations}
A \emph{qualified constrained interpretation} (or {qc-interpreta\-tion})
is a set $\I$ of defined observable qc-atoms closed under $(\qdom, \cdom)$-entailment, i.e. $\varphi \in \I$ and
$\varphi \entail{\qdom,\cdom} \varphi'$ implies $\varphi' \in \I$.
An observable qc-atom $\varphi$ is called {\em valid} in the qc-interpretation $\I$ (in symbols, $\I \isqchlrdc \varphi$)
iff some of the following cases holds:
{\sf (a)} $\varphi$ is a defined qc-atom and $\varphi \in \I$; 
or {\sf (b)} $\varphi$ is an equational qc-atom $\cqat{(t == s)}{d}{\Pi}$ and $t \approx_{d, \Pi} s$;
or {\sf (c)} $\varphi$  is a primitive qc-atom $\cqat{\kappa}{d}{\Pi}$ and $\Pi \model{\cdom} \kappa$. \mathproofbox
\end{defn}

Note that a given interpretation $\I$ can include several observables $\cqat{A}{d_i}{\Pi}$ for the same (possibly open) atom $A$, but is not required to include one ``optimal'' observable $\cqat{A}{d}{\Pi}$ with $d$ computed as the $lub$ of all $d_i$.
By contrast, the other related works discussed in the Introduction view program interpretations 
as mappings $\I$ from the ground Herbrand base into some set of lattice elements (the real interval $[0,1]$ in many cases).
In such interpretations, each ground atom $A$ has attached one single lattice element $d = \I(A)$ intended as ``the optimal qualification'' for $A$.
Our view of interpretations is closer to the expected operational behavior of goal solving systems and can be used to characterize the validity of solutions computed by such systems, as we will see in Subsection \ref{sec:sqclp:goals}.

It can be proved that $\I \isqchlrdc \varphi$ implies $\I \isqchlrdc \varphi'$ for any
$\varphi'$ such that $\varphi \entail{\qdom,\cdom} \varphi'$ (so-called {\em entailment property for interpretations}).
The notions of model and semantic consequence are defined below.


\begin{defn}[Models and semantic consequence]
\label{defn:models} 
Let a $\sqclp{\simrel}{\qdom}{\cdom}$-program $\Prog$ and
an observable qc-atom $\varphi : \cqat{p'(\ntup{t'}{n})}{d}{\Pi}$ be given. 
$\varphi$  is an {\em immediate consequence} of a 
qc-interpretation $\I$ via a program rule $(R_l :  p(\ntup{t}{n}) \qgets{\alpha} \qat{B_1}{w_1}, \ldots,$ $\qat{B_m}{w_m}) \in \Prog$ iff
there exist a $\cdom$-substitution $\theta$ and a choice of qualification values $d_0, d_1, \ldots,$ $d_n, e_1, \ldots, e_m \in \aqdom$
such that:
     \begin{enumerate}
        \item[(a)]
        $\simrel(p',p) = d_0$,
        \item[(b)]
        $\I \isqchlrdc \cqat{(t'_i == t_i\theta)}{d_i}{\Pi}$ (i.e. $t'_i \approx_{d_i,\Pi} t_i\theta$) for $i = 1 \ldots n$,
        \item[(c)]
        $\I \isqchlrdc \cqat{B_j\theta}{e_j}{\Pi}$ with $e_j \dgeq^? w_j$ for $j = 1 \ldots m$,
        \item[(d)]
        $d \dleq \bigsqcap_{i = 0}^{n}d_i \sqcap \alpha \circ \bigsqcap_{j = 1}^m e_j$.
      \end{enumerate}
 Note that  the qualification value $d$ attached to $\varphi$  is limited by two kinds of upper bounds:
$d_i ~ (0 \le i \le n)$, i.e. the $\simrel$-proximity between $p'(\ntup{t'}{n})$ and the head
of $R_l\theta$; and $\alpha \circ e_j ~ (1 \le j \le m)$, i.e.  the qualification values of the
atoms in the body of $R_l\theta$ attenuated w.r.t. $R_l$'s attenuation factor $\alpha$.
Now we can define:
\begin{enumerate}
 \item
  $\I$ is a \emph{model} of $R_l \in \Prog$
  (in symbols, $\I \model{\simrel,\qdom,\cdom} R_l$)
  iff every defined observable qc-atom $\varphi$ that is an immediate consequence of $\I$ via $R_l$ verifies $\varphi \in \I$.
  And  $\I$ is a \emph{model} of $\Prog$ (in symbols, $\I \model{\simrel,\qdom,\cdom} \Prog$)
  iff $\I$ is a model of each $R_l \in \Prog$.
 \item
  $\varphi$ is a \emph{semantic consequence} of $\Prog$ (in symbols, $\Prog \model{\simrel,\qdom,\cdom} \varphi$)
  iff $\I \isqchlrdc \varphi$ for every qc-interpretation $\I$ such that $\I \model{\simrel,\qdom,\cdom} \Prog$. \mathproofbox
\end{enumerate}
\end{defn}

The next example may serve as a concrete illustration:

\begin{exmp}[Models and semantic consequence]
\label{exmp:semantic-consequence} Recall the $\sqclp{\simrel}{\U}{\rdom}$-program $\Prog$ from
Example \ref{exmp:running} and the qc-atoms $\varphi_1$ and $\varphi_2$ from Example
\ref{exmp:qc-atoms}. Assume an arbitrary model $\I \model{\simrel,\U,\rdom} \Prog$. Then:

\noindent --- (1)
Note that the atom underlying $\varphi_1$ is $q(X,c'(Y))$, and the head atom of $R_1$ is $q(X,c(X))$.
Since $\simrel(c,c') = 0.9$ and $\Pi \model{\cdom} X == Y$, $\varphi_1$ can be obtained as an immediate consequence of $\I$ via $R_1$ using $\theta = \varepsilon$.
Therefore, $\varphi_1 \in \I$ and $\Prog \model{\simrel,\U,\rdom} \varphi_1$.

\noindent --- (2)
Consider $\theta = \{Y \mapsto c'(Y)\}$.
Note that $p'(c'(Y),c(X))$ is the atom underlying $\varphi_2$, and the head atom of $R_2\theta$ is $p(c(X),c'(Y))$.
Moreover, $\varphi_1 \in \I$ due to the previous item and the atom $q(X,c'(Y))$ underlying $\varphi_1$ is the same as the atom in the body of $R_2\theta$.
These facts together with $\simrel(p,p') = 0.8$, $\simrel(c,c') = 0.9$  and $\Pi \model{\cdom} X == Y$ allow to obtain $\varphi_2$ as an immediate consequence of $\I$ via $R_2$.
Therefore, $\varphi_2 \in \I$ and $\Prog \model{\simrel,\U,\rdom} \varphi_2$. \mathproofbox
\end{exmp}


\subsection{A Fixpoint Semantics}
\label{sec:sqclp:fixpointsem}

As for other declarative languages, one can use immediate consequence operators to characterize the
models and least models of  a given $\sqclp{\simrel}{\qdom}{\cdom}$-program $\Prog$. 
We start by considering the complete lattice $\Intdc$ of all qc-interpretations
partially ordered by set inclusion, with bottom element $\ibot = \emptyset$ and top element
$\itop = \{\varphi \mid \varphi$ is a defined observable qc-atom\}. For any subset $I \subseteq \Intdc$
one gets the greatest lower bound $\infi I = \inter_{\I \in I} \I$ and the least upper bound  $\supr I = \union_{\I \in I} \I$.
Next we define an \emph{interpretation transformer} $\Tp : \Intdc \to \Intdc$, intended to compute the immediate
consequences obtained from a given qc-interpretation via the program rules belonging to $\Prog$,
and defined as
$$\Tp(\I) \eqdef \{ \varphi \mid \varphi \mbox{ is an immediate consequence of } \I \mbox{ via some } R_l \in \Prog\}$$
where immediate consequences are computed as explained in Definition \ref{defn:models}. The following example
illustrates the workings of $\Tp$.

\begin{exmp}[Interpretation transformer in action]
\label{exmp:tp} Recall again the $\sqclp{\simrel}{\U}{\rdom}$-program $\Prog$ from Example
\ref{exmp:running} and the defined observable qc-atoms $\varphi_1$ and $\varphi_2$
from Example \ref{exmp:qc-atoms}. Then:
{\sf (1)}
The arguments given in Example \ref{exmp:semantic-consequence}(1)
can be easily reused to show that $\varphi_1$ is an immediate consequence of $\ibot$ via $R_1$. 
Therefore, $\varphi_1 \in \Tp(\ibot)$.
{\sf (2)}
The arguments given in Example \ref{exmp:semantic-consequence}(2)
can be easily reused to show that $\varphi_2$ is an immediate consequence of $\I$ via $R_2$,
provided that $\varphi_1 \in \I$. Therefore, $\varphi_2 \in \Tp(\Tp(\ibot))$. \mathproofbox
\end{exmp}

The next proposition states the main properties of interpretation transformers.

\begin{prop}[Properties of interpretation transformers]
\label{prop:tp-properties}
For any $\sqclp{\simrel}{\qdom}{\cdom}$-program $\Prog$,
$\Tp$ is a well defined mapping, i.e. for all $\I \in \Intdc$ one has  $\Tp(\I) \in \Intdc$.
Moreover, $\Tp$ is monotonic and continuous and its pre-fixpoints are the models of $\Prog$, i.e.
for all $\I \in \Intdc$ one has $\I \model{\simrel,\qdom,\cdom} \Prog \Longleftrightarrow \Tp(\I) \subseteq \I$. \mathproofbox
\end{prop}

As an immediate consequence one can prove the theorem below, that is the main result in this subsection.

\begin{thm}[Fixpoint characterization of least program models]
\label{thm:tp-leastmodel} Every $\sqclp{\simrel}{\qdom}{\cdom}$-program $\Prog$ has a \emph{least
model} $\Mp$, smaller than any other model of $\Prog$ w.r.t. the set inclusion ordering of the
interpretation lattice $\Intdc$. Moreover, $\Mp$ can be characterized as \emph{least fixpoint} of $\Tp$ as follows:
$$\Mp = l\!f\!p(\Tp) = \union_{k\in\NAT} \Tp{\uparrow^k}(\ibot) \enspace . \mathproofbox$$
\end{thm}

\subsection{An equivalent Proof-theoretic Semantics}
\label{sec:sqclp:sqchl}


In order to give a logical view of program semantics and an alternative characterization of least
program models, we define the \emph{Proximity-based Qualified Constrained Horn Logic}
$\SQCHL(\simrel,\qdom,\cdom)$ as a formal inference system consisting of the three inference rules
displayed in Figure \ref{fig:sqchl}.

\begin{figure}[h]
  \figrule
  \centering
  \vspace{2mm}
  \begin{tabular}{l}
  \textbf{SQDA} ~ $\displaystyle\frac
    {~ (~ \cqat{(t'_i == t_i\theta)}{d_i}{\Pi} ~)_{i=1 \ldots n} \quad (~ \cqat{B_j\theta}{e_j}{\Pi} ~)_{j=1 \ldots m} ~}
    {\cqat{p'(\ntup{t'}{n})}{d}{\Pi}}$ \\ \\
    \hspace{13mm} if $(p(\ntup{t}{n}) \qgets{\alpha} \qat{B_1}{w_1}, \ldots, \qat{B_m}{w_m}) \in \Prog$\!,~
        $\theta$ subst.,~ $\simrel(p',p) = d_0 \neq \bt$, \\
    \hspace{13mm} $e_j \dgeq^? w_j ~ (1 \le j \le m)$ and 
        $d \dleq \bigsqcap_{i = 0}^{n}d_i \sqcap \alpha \circ \bigsqcap_{j = 1}^m e_j$. \\
  \\
  \textbf{SQEA} ~ $\displaystyle\frac
    {}
    {\quad \cqat{(t == s)}{d}{\Pi} \quad}$ ~
  if $t \approx_{d, \Pi} s$. 
  
  ~~ \textbf{SQPA} ~ $\displaystyle\frac
    {}
    {\quad \cqat{\kappa}{d}{\Pi} \quad}$ ~
  if $\Pi \model{\cdom} \kappa$. \\
  \end{tabular}
  \vspace{2mm}
  \caption{Proximity-based Qualified Constrained Horn Logic}
  \label{fig:sqchl}
  \figrule
  \vspace{-3mm}
\end{figure}


Rule \textbf{SQDA} formalizes an extension of  the classical \emph{Modus Ponens}
inference allowing to infer a defined qc-atom $\cqat{p'(\ntup{t'}{n})}{d}{\Pi}$ by means of an
instantiated clause with head $p(\ntup{t}{n})\theta$ and body atoms
$\qat{B_j\theta}{w_j}$. The $n$ premises $\cqat{(t'_i == t_i\theta)}{d_i}{\Pi}$ combined with the
side condition $\simrel(p',p) = d_0 \neq \bt$  ensure the ``equality'' between
$p'(\ntup{t'}{n})$ and $p(\ntup{t}{n})\theta$ modulo $\simrel$;
the $m$ premises $\cqat{B_j\theta}{e_j}{\Pi}$ require to prove the body atoms;
and the side conditions $e_j \dgeq^? w_j$ and $d \dleq \bigsqcap_{i =
0}^{n}d_i \sqcap \alpha \circ \bigsqcap_{j = 1}^m e_j$ check the threshold conditions of the body
atoms and impose the proper relationships between the qualification
value $d$ attached to the conclusion and the qualification values $d_i$ and $e_j$ attached to the premises.
Rule \textbf{SQEA} is designed to work with constraint-based term proximity  in the sense
of Definition \ref{defn:Pi-prox}, inferring $\cqat{(t  == s)}{d}{\Pi}$ just in the case
that $t \approx_{d, \Pi} s$ holds. Rule \textbf{SQPA} infers primitive qc-atoms $\cqat{\kappa}{d}{\Pi}$
for an arbitrary $d \in \aqdom$, provided that $\Pi \model{\cdom} \kappa$ holds.

We will write $\Prog \sqchlrdc \varphi$ to indicate that $\varphi$ can be deduced from $\Prog$ in 
$\SQCHL(\simrel,$ $\qdom,\cdom)$, and $\Prog \sqchlrdcn{k} \varphi$ in the case that the deduction can be performed with exactly $k$ {\bf SQDA} inference steps.
As usual in formal inference systems, $\SQCHL(\simrel,\qdom,\cdom)$ proofs can be represented as {\em proof trees} whose nodes correspond to qc-atoms, each node being inferred from its children by means of some $\SQCHL(\simrel,\qdom,\cdom)$ inference step.
The next example  shows a simple $\SQCHL(\simrel,\U,\rdom)$ proof tree.

\begin{exmp}[$\SQCHL(\simrel,\qdom,\cdom)$ proof tree]
\label{exmp:sqchl-inference} Recall the proximity relation $\simrel$ and the program $\Prog$ from
our running example \ref{exmp:running} and the observable qc-statement $\varphi_1 =
\cqat{q(X,c'(Y))}{0.9}{\Pi}$ already known from Example \ref{exmp:qc-atoms}. A
$\SQCHL(\simrel,\U,\rdom)$ proof tree witnessing $\Prog \sqchln{\simrel}{\U}{\rdom}{1} \varphi_1$ can
be displayed as follows:
$$
\displaystyle\frac {\quad
  \displaystyle\frac{}{\cqat{(X == X)}{1.0}{\Pi}} ~ {\bf\scriptstyle SQEA (2)} \quad
  \displaystyle\frac{}{\cqat{(c'(Y) == c(X))}{0.9}{\Pi}} ~ {\bf\scriptstyle SQEA (3)}
~} {\cqat{q(X,c'(Y))}{0.9}{\Pi}} ~ {\bf\scriptstyle SQDA (1)}
$$
Where:
step {\sf (1)} uses $R_1 = q(X,$ $c(X)) \qgets{1.0}$
instantiated by the empty substitution (note that $0.9 \le \mbox{min}\{1.0,0.9\}$);
step {\sf (2)} uses $X \approx_{1.0, \Pi} X$, trivially true; and
step {\sf (3)} uses $c'(Y) \approx_{0.9, \Pi} c(X)$, true due to $\simrel$(c,c') = 0.9 and
$\Pi \model{\rdom} X == Y$. \mathproofbox
\end{exmp}

It can be proved that $\Prog \sqchlrdc \varphi$ implies $\Prog \sqchlrdc \varphi'$ for any
$\varphi'$ such that $\varphi \entail{\qdom,\cdom} \varphi'$ (so-called {\em entailment property for programs}).
Moreover, if $\varphi$ is either equational or primitive, then
$\Prog  \sqchlrdcn{0} \varphi \Longleftrightarrow \Prog \sqchlrdc \varphi \Longleftrightarrow \I \isqchlrdc \varphi$
for any program $\Prog$ and any qc-interpretation $\I$.
The following theorem is the main result in this subsection.

\begin{thm}[Logical characterization of least program models]
\label{thm:SQCHL-leastmodel} For any $\sqclp{\simrel}{\qdom}{\cdom}$-program $\Prog$, its least
model can be characterized as: $$\Mp = \{\varphi \mid \varphi \mbox{ is an observable defined
qc-atom and }\Prog \sqchlrdc \varphi\}. \enspace \mathproofbox$$
\end{thm}

As an easy consequence of the previous theorem we can prove:

\begin{cor}[$\SQCHL(\simrel,\qdom,\cdom)$ is sound and complete]
\label{cor:correctness} For any $\sqclp{\simrel}{\qdom}{\cdom}$-program $\Prog$ and any observable
qc-atom $\varphi$ one has:
\begin{enumerate}
  \item $\Prog \sqchlrdc \varphi \Longleftrightarrow \Prog \model{\simrel,\qdom,\cdom} \varphi \Longleftrightarrow \Mp \isqchlrdc \varphi$.
  \item $\Prog \sqchlrdc \varphi \Longrightarrow \Prog \model{\simrel,\qdom,\cdom} \varphi$ ({\em soundness}).
  \item $\Prog \model{\simrel,\qdom,\cdom} \varphi \Longrightarrow \Prog \sqchlrdc \varphi$ ({\em completeness}). \mathproofbox
\end{enumerate}
\end{cor}

\subsection{Goals and solutions}
\label{sec:sqclp:goals}

In order to build goals for $\sqclp{\simrel}{\qdom}{\cdom}$-programs, we assume a countably
infinite set $\War$ of so-called {\em qualification variables} $W$\!. Goals for a given program $\Prog$ have the form:
$$G ~:~ \qat{A_1}{W_1},~ \ldots,~ \qat{A_m}{W_m} \sep W_1 \dgeq^? \!\beta_1,~ \ldots,~ W_m \dgeq^? \!\beta_m$$
abbreviated as $(\qat{A_i}{W_i}, W_i \dgeq^? \!\beta_i)_{i = 1 \ldots m}$ with annotated atoms
$\qat{A_i}{W_i}$ (where the qualification variables $W_i \in \War$ are pairwise different) and threshold conditions 
$W_i \dgeq^? \!\beta_i$ with $\beta_i \in \bqdom$. The notations ? and $\dgeq^?$ have been explained in Subsection \ref{sec:sqclp:programs}.

The proof-theoretical semantics developed in Subsection \ref{sec:sqclp:sqchl} allows to characterize
 {\em goal solutions} in a natural and declarative way by means of the following definition:
the set  of solutions of a goal $G$ w.r.t. program $\Prog$ is noted $\Sol{\Prog}{G}$ and 
consists of all triples $\langle \sigma, \mu, \Pi \rangle$ such that
$\sigma$ is a $\cdom$-substitution (not required to be ground), $\mu : \{W_1, \ldots, W_m\} \to \aqdom$, 
$\Pi$ is a satisfiable and finite set of atomic $\cdom$-constraints
and the following two conditions hold for all $i = 1 \ldots m$:
$W_i\mu = d_i \dgeq^? \!\beta_i$ and $\Prog \sqchlrdc \cqat{A_i\sigma}{W_i\mu}{\Pi}$. 
Although operational semantics is not investigated in this paper,
{\em computed answers} obtained by means of a correct goal solving system for $\sqclp{\simrel}{\qdom}{\cdom}$
are expected to be valid solutions in this sense.

For instance,  $G : \qat{\texttt{goodWork(X)}}{\texttt{W}}  \sep \texttt{W} \dgeq \texttt{(0.55,30)}$
is a goal for the program fragment  $\Prog$ shown  in Figure \ref{fig:sample},
and the arguments given near the beginning of Subsection \ref{sec:sqclp:programs} can be formalized to prove that 
$\langle \{\texttt{X} \mapsto \texttt{king\_liar}\}, \{\texttt{W} \mapsto \texttt{(0.6,5)}\}, \emptyset \rangle \in \mbox{Sol}_\Prog(G)$.

As an additional example involving constraints, recall the $\sqclp{\simrel}{\U}{\rdom}$ program $\Prog$ presented in Example \ref{exmp:running} and consider the goal $G : \qat{q(X,Z)}{W} \sep W \geq 0.8$ for $\Prog$.
Then $\langle \sigma, \mu, \Pi \rangle \in \Sol{\Prog}{G}$,
where  $\sigma = \{Z \mapsto c'(Y)\}$, 
$\mu = \{W \mapsto 0.9\}$ and $\Pi = \{cp_{>}(X,1.0)$, 
$op_{+}(A,A,X)$, $op_{\times}(2.0,A,Y)\}$.
Note that $W\!\mu = 0.9 \geq 0.8$ and $\Prog \vdash_{\simrel,\U,\rdom} \cqat{q(X,Z)\sigma}{0.9}{\Pi}$ 
is known from Example \ref{exmp:sqchl-inference}.

\section{Conclusions}
\label{sec:conclusions}


We have extended the classical CLP scheme to a new scheme SQCLP whose instances  $\sqclp{\simrel}{\qdom}{\cdom}$ are
parameterized by a proximity relation $\simrel$, a qualification domain $\qdom$ and a constraint domain $\cdom$.
In addition to the known features of CLP programming, the new scheme offers extra facilities for dealing with expert knowledge representation and flexible query answering.
Inspired by the observable CLP semantics in \cite{GDL95},
we have presented a declarative semantics for SQCLP that 
provides fixpoint and proof-theoretical characterizations of least program models
as well as an implementation-independent notion of goal solutions.

SQCLP is a quite general scheme. Different partial instantiations of its three parameters lead to 
more particular schemes, most of which can be placed in close correspondence to previous proposals.
The items below present seven particularizations, along with some comments which make use of the following terminology:
a SCQLP program is called {\em threshold-free} in case that all its clauses use only `?' as threshold value;
{\em attenuation-free}  in case that all its clauses use only $\tp$ as attenuation value;
and {\em constraint-free} in case that no constraints occur in clause bodies.

\begin{enumerate}
\item 
By definition, QCLP has instances $\mbox{QCLP}(\qdom,\cdom) \eqdef \mbox{SQCLP}(\sid,\qdom,\cdom)$
where $\sid$ is the {\em identity} proximity relation.
The {\em quantitative} CLP scheme proposed in \cite{Rie98phd} 
can be understood as a further particularization of QCLP 
that works with threshold-free $\mbox{QCLP}(\U,\cdom)$ programs,
where $\U$ is the qualification domain of uncertainty values (see Subsection \ref{sec:cbasis:qdoms}).
\item 
By definition, SQLP has instances $\mbox{SQLP}(\simrel,\qdom) \eqdef \mbox{SQCLP}(\simrel,\qdom,\rdom)$
where $\rdom$ is the real constraint domain (see Subsection \ref{sec:cbasis:cdoms}).
The scheme with the same name originally proposed in \cite{CRR08} 
can be understood as a restricted form of the present formulation;
it worked with  threshold-free and constraint-free $\mbox{SQLP}(\simrel,\qdom)$ programs
and it restricted the choice of the $\simrel$ parameter to transitive proximity (i.e. similarity)  relations. 
\item 
By definition, SCLP\footnote{Not to be confused with SCLP in the sense of \cite{BMR01}, discussed below.} has instances $\mbox{SCLP}(\simrel,\cdom) \eqdef \mbox{SQCLP}(\simrel,\B,\cdom)$
where $\B$ is the qualification domain of classical boolean values  (see Subsection \ref{sec:cbasis:qdoms}).
Due to the fixed parameter choice $\qdom = \B$,
both attenuation values and threshold values become useless,  
and each choice of $\simrel$ must necessarily represent a crisp reflexive and symmetric relation.
Therefore, this new scheme is not so interesting from the viewpoint of uncertain and qualified reasoning.
\item 
By definition, QLP has instances $\mbox{QLP}(\qdom) \eqdef \mbox{SQCLP}(\sid,\qdom,\rdom)$.
The original scheme with the same name proposed  in \cite{RR08} 
can be understood as a restricted form of the present formulation;
it worked with  threshold-free and constraint-free $\mbox{QLP}(\qdom)$ programs.
\item 
By definition, SLP has instances $\mbox{SLP}(\simrel) \eqdef  \mbox{SQCLP}(\simrel,\U,\rdom)$.
The pure fragment of \textsf{Bousi}$\sim$\textsf{Prolog} \cite{JR09}
can be understood as a restricted form of SLP in the present formulation;
it works with threshold-free, attenuation-free and constraint-free $\mbox{SLP}(\simrel)$ programs.
Moreover, restricting the choice of $\simrel$ to similarity relations leads to 
SLP in the sense of \cite{Ses02} and related papers.
\item 
The CLP scheme can be defined  by  instances $\mbox{CLP}(\cdom) \eqdef \mbox{SQCLP}(\sid,\B,\cdom)$. 
Both attenuation values and threshold values are useless in CLP programs,  due to the fixed parameter choice $\qdom = \B$.
\item 
Finally, the pure LP paradigm can be defined as $\mbox{LP} \eqdef  \mbox{SQCLP}(\sid,\B,\hdom)$
where $\hdom$ is the {\em Herbrand} constraint domain.
Again, attenuation values and threshold values are useless in LP due to the fixed parameter choice $\qdom = \B$.
\end{enumerate}

In all the previous items, the schemes obtained by partial instantiation inherit the declarative semantics from  
SQCLP, using sets of observables of the form $\cqat{A}{d}{\Pi}$ as interpretations. 
Similar semantic approaches were used in our previous papers \cite{RR08,CRR08}, 
except that $\Pi$ and equations were absent due to the lack of CLP features.
The other related works discussed in the Introduction view program interpretations 
as mappings $\I$ from the ground Herbrand base into some set of lattice elements (the real interval $[0,1]$ in many cases), as already discussed in the explanations following Definition \ref{defn:interpretations}.

As seen in Subsection \ref{sec:sqclp:goals}, SQCLP's semantics enables a declarative characterization of valid goal solutions. This fact is relevant for modeling the expected behavior of goal solving devices and reasoning about their correctness.
Moreover, the relations $\approx_{\lambda, \Pi}$ introduced for the first time in the present paper (see Definition \ref{defn:Pi-prox}) allow to specify the semantic role of $\simrel$ in a constraint-based framework, with less technical overhead than in previous related approaches.


A related work not mentioned in items above is semiring-based CLP \cite{BMR01}, a scheme with instances SCLP(S) parameterized by a semiring $\mbox{S} = \langle A, +, \times, \mathbf{0}, \mathbf{1} \rangle$ whose elements are used to represent consistency levels in soft constraint solving.
The semirings used in this approach can be equipped with a lattice structure whose {\em lub} operation is always $+$, but whose {\em glb} operation may be different from $\times$.
On the other hand, our qualification domains are defined as lattices with an additional attenuation operation $\circ$.
It turns out that the kind of semirings used in SCLP(S) correspond to qualification domains only in some cases.
Moreover, $\times$ is used in SCLP(S) to interpret logical conjunction in clause bodies and goals, while the {\em glb} operation is used in instances $\sqclp{\simrel}{\qdom}{\cdom}$ for the same purpose. For this reason, even if $\qdom$ is ``equivalent'' to S, $\sqclp{\simrel}{\qdom}{\cdom}$ cannot be naturally used to express SCLP(S) in the case that $\times$ is not the {\em glb}.
Assuming that $\qdom$ is ``equivalent'' to S and that $\times$ behaves as the {\em glb} in S, program clauses in SCLP(S) can be viewed as a particular case of program clauses in $\sqclp{\simrel}{\qdom}{\cdom}$ which use an attenuation factor different from $\tp$ only for facts.
Other relevant differences between $\sqclp{\simrel}{\qdom}{\cdom}$ and SCLP(S) can be explained by comparing the parameters.
As said before $\qdom$ may be ``equivalent'' to S in some cases, but $\simrel$ is absent and $\cdom$ is not made explicit in SCLP(S).
Seemingly, the intended use of SCLP(S) is related to finite domain constraints and no parametrically given constraint domain is provided.


In the future we plan to implement some SQCLP instances by extending the 
semantically correct program transformation techniques from \cite{CRR08},
and to investigate applications which can profit from flexible query answering in the line of \cite{CDG+09} and other related papers.
Other interesting lines of future work include:
a)  extension of the qualified SLD resolution presented in \cite{RR08} 
to a SQCLP goal solving procedure able to work with constraints and proximity relations; and
b) extension of the QCFLP scheme in \cite{CRR09} to work with proximity relations and higher-order functions.

\section*{Acknowledgements}

The authors are thankful to the anonymous referees for constructive remarks and suggestions which helped to improve the presentation.
They are also thankful to Rafael Caballero for useful discussions on the paper's topics
and to Jes\'us Almendros for pointing to bibliographic references in the area of flexible query answering.


\end{document}